\documentclass[conference,a4paper]{APSIPA2021}
\usepackage{amsmath}
\usepackage{graphicx}
\usepackage{multirow}
\usepackage{amsfonts}
\usepackage{amssymb}
\usepackage{threeparttable}
\usepackage{pgfplots} 
\pgfplotsset{width=9cm, compat=1.6}  
\usepackage[backend=biber,style=ieee,]{biblatex}
\usepackage{geometry}
\usepackage{fancyhdr}

\fancypagestyle{firststyle}{
  \fancyhf{}
}

\begin{document}

\title{RobustL2S: Speaker-Specific Lip-to-Speech Synthesis exploiting Self-Supervised Representations}

\author{
\authorblockN{
Neha Sahipjohn\authorrefmark{1} \mbox{ }
Neil Shah\authorrefmark{1}\authorrefmark{2} \mbox{ }
Vishal Tambrahalli\authorrefmark{1} \mbox{ }
Vineet Gandhi\authorrefmark{1} \mbox{ }
}

\authorblockA{
\authorrefmark{1}
CVIT, Kohli Centre for Intelligent Systems, IIIT Hyderabad, India}

 \authorrefmark{2}
 TCS Research, Pune, India \\
 }


\maketitle
\thispagestyle{firststyle}
\pagestyle{fancy}
\newcommand{\ourmodel}{RobustL2S}
\newcommand{\lts}{Lip-to-Speech}
\newcommand{\nar}{non-autoregressive}
\newcommand{\sts}{Seq2Seq}
\begin{abstract}
 Significant progress has been made in speaker-dependent Lip-to-Speech synthesis, which aims to generate speech from silent videos of talking faces. Current state-of-the-art approaches primarily employ non-autoregressive sequence-to-sequence architectures to directly predict mel-spectrograms or audio waveforms from lip representations. We hypothesize that the direct mel-prediction hampers training/model efficiency due to the entanglement of speech content with ambient information and speaker characteristics. To this end, we propose \ourmodel, a modularized framework for Lip-to-Speech synthesis. First, a non-autoregressive sequence-to-sequence model maps self-supervised visual features to a representation of disentangled speech content. A vocoder then converts the speech features into raw waveforms. Extensive evaluations confirm the effectiveness of our setup, achieving state-of-the-art performance on the unconstrained Lip2Wav dataset and the constrained GRID and TCD-TIMIT datasets. Speech samples from \ourmodel~can be found at \textcolor{blue}{\url{https://neha-sherin.github.io/RobustL2S/}}
  




\end{abstract}

\section{Introduction}
Understanding lip movements offers a distinct advantage in situations where auditory cues are unavailable. It proves particularly valuable for individuals with hearing impairments, speech disorders and aids in speech rehabilitation by providing visual feedback \cite{su2023liplearner}. The synthesis of accurate speech from lip movements can assist in tasks such as movie dubbing \cite{cong2023learning}, language learning, forensic investigations \cite{el2023developing}, video conferencing in noisy conditions, voice inpainting \cite{zhou2019vision} or giving artificial voice to people who cannot produce intelligible sound. 



The problem of \lts~synthesis is inherently ill-posed because a sequence of lip movements can correspond to multiple possible speech utterances~\cite{prajwal2020learning}. Additional challenges arise from factors such as head pose movements, non-verbal facial expressions, variations in capture quality, and ambient noise, which further complicate the problem. Reliance on contextual information, such as environment, place, topic, etc., can help alleviate the Lipreading challenges ~\cite{sumby1954visual,bernstein2022lipreading}; however, such information may not always be available.



Most existing approaches constitute an encoder-decoder architecture; the encoder maps the lip sequence to intermediate representations, which are then directly decoded into mel-spectrograms. The major drawback of this approach is that apart from speech content, the decoder is also forced to predict the time-varying speaker and ambient noise characteristics present in the ground truth Mel. We hypothesize that this dependence hurts the model's performance in terms of speech intelligibility, reducing its usability for various downstream applications \cite{choi2023intelligible}. Our work addresses these limitations by taking a modularized approach, exploiting the advances in Self-Supervised Learning (SSL) in audio and audio-visual scenarios.

\begin{figure}[t]
  \centering
\includegraphics[width=\linewidth]{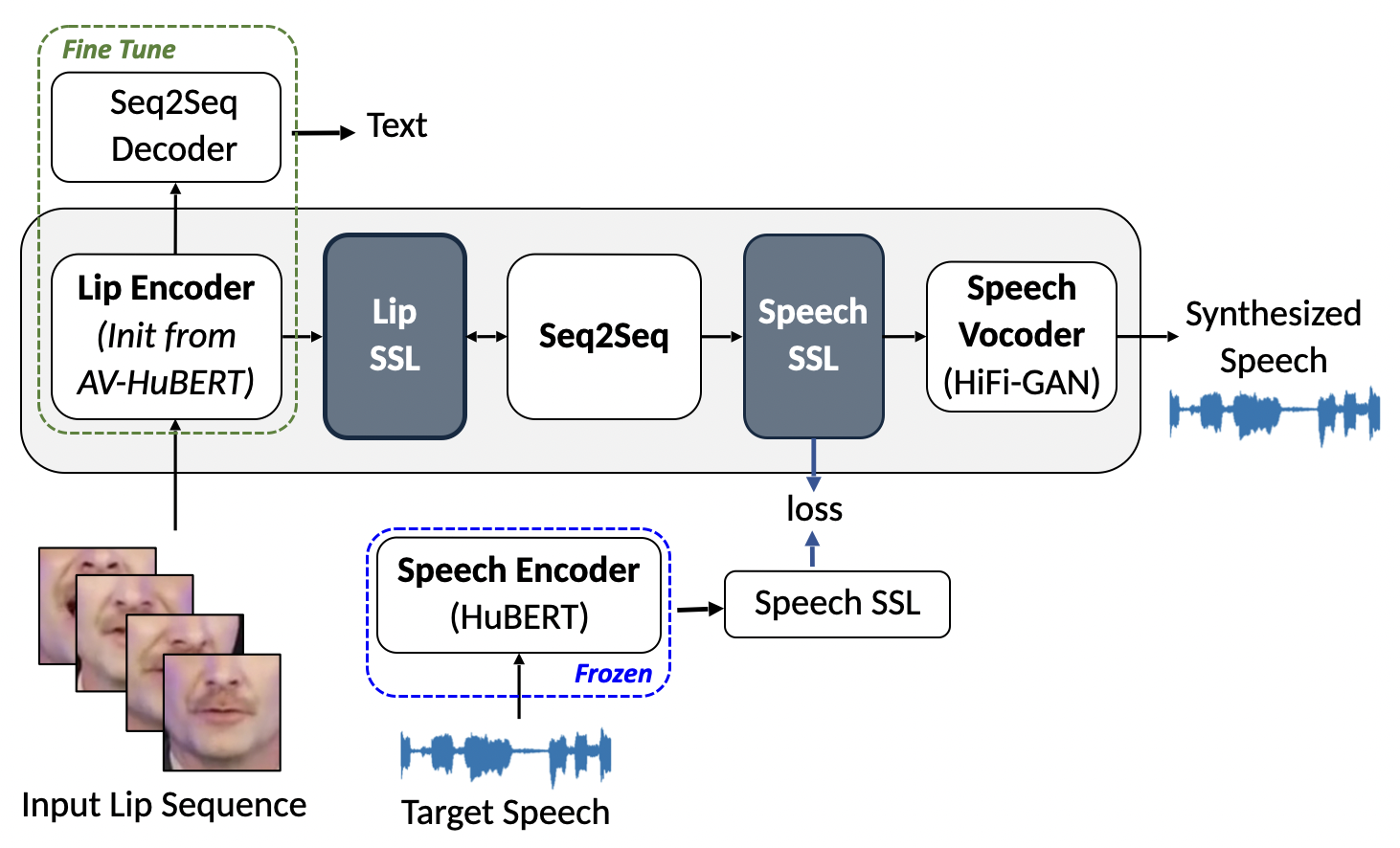}
\caption{The proposed \ourmodel~model utilizes lip encoder and speech encoder to extract SSL representations from lip sequences and their corresponding speech. A \sts~model maps the lip representations to speech representations, which are then decoded to synthesize speech.} 
  \label{fig:teaser}
\end{figure}


Fig.~\ref{fig:teaser} illustrates the proposed \ourmodel~framework. In contrast to direct mel prediction from lip features, we take a two-staged approach. The first step extracts SSL representations of lip sequences and maps them to corresponding speech SSL representations using a sequence-to-sequence (\sts)~model. The key idea is to use speech embeddings that disentangle the content from the speaker and ambient information. The second stage maps the content-rich speech embeddings to raw speech using a speaker-conditioned vocoder. The proposed \ourmodel~ framework simplifies training and brings robustness to variations in head-pose, ambient noise, and time-varying speaker characteristics, leading to significant gains in speech intelligibility. To validate the efficacy of our approach, we perform comprehensive experiments on GRID \cite{cooke2006audio}, TCD-TIMIT~\cite{harte2015tcd} and Lip2Wav \cite{prajwal2020learning} datasets. The quantitative measures and MOS scores show that the synthesized speech generated by our method accurately represents the intended content and improves on the intelligibility/naturalness compared to current state-of-the-art methods \cite{wang2022fastlts,wang2022vcvts} on all three datasets.







More formally, our work makes the following contributions: (1) We propose a novel modularized framework for \lts~synthesis exploiting self-supervised embeddings for both lip and speech sequences (2) A \sts~network for cross-modal knowledge transfer to map lip SSL representations to speech SSL representations; and 
(3) Thorough experimental results demonstrating that \ourmodel~is capable of synthesizing high-quality speech, achieving state-of-the-art results in objective and subjective evaluation without requiring additional data augmentation~\cite{wang2022fastlts}.

\section{Related Work}
\subsection{\lts~synthesis}
\subsubsection{Constrained \lts~synthesis}
Constrained lip-to-speech synthesis tackles speech generation from videos with limited vocabulary and minimal head movement~\cite{cooke2006audio,harte2015tcd}. Ephrat et al.\cite{ephrat2017vid2speech} introduced a CNN-based approach to predict Linear Predictive Coding features from silent talking videos. They later enhanced their model to a two-tower CNN-based encoder-decoder architecture \cite{ephrat2017improved}, encoding raw frames and optical flows separately. \cite{akbari2018lip2audspec} propose a combination of an autoencoder for extracting bottleneck features from audio spectrograms and a lipreading network comprising CNN, LSTM, and fully connected layers, for visual feature extraction. On the other hand, \cite{yadav2021speech} utilize a stochastic modeling approach employing a variational autoencoder. Other methods \cite{vougioukas2019video, mira2022end, kim2021lip,mira2022svts} employ GANs to synthesize speech from video frames. \cite{qu2022lipsound2} train an attention-based encoder-decoder model to reconstruct speech from silent facial movement sequences without human annotations. \cite{kim2023lip} performs multi-modal supervision, using text and audio, to complement the insufficient word representations to reconstruct speech with correct contents from the input lip movements. \cite{sheng2023zero} achieves zero-shot \lts~synthesis using variational autoencoder to disentangle speaker and content information and a face identity encoder for unseen speakers. 


\subsubsection{Unconstrained \lts~synthesis}
Unconstrained lip-to-speech synthesis focuses on generating speech from real-world videos of individuals talking. This approach considers videos with an extensive vocabulary and significant head movements. Seminal work by Prajwal et \textit{al}.~\cite{prajwal2020learning} introduced a 3D-convolution plus autoregressive \sts~model, adapted from Tacotron2~\cite{shen2018natural} for single speaker \lts~synthesis. Their model generates mel-spectrograms based on the input video frames. In contrast, \cite{he2022flow} employed a \nar~architecture to expedite the inference process. It uses 3D-convolution blocks, a transformer condition module, and a Glow decoder module \cite{kingma2018glow} for refining the mel-spectrograms. \cite{varshney2022learning} expanded on \cite{prajwal2020learning} by training a transformer model to learn a joint latent distribution for speech generation. The VV-Memory architecture \cite{hong2021speech} combines audio and visual information using a key-value memory structure to enable video-to-speech reconstruction and speaker-independent speech retrieval. A recent work by \cite{wang2022fastlts} proposed an end-to-end \nar~ transformer for synthesizing speech directly from unconstrained talking videos. Their approach involves a visual encoder, an acoustic decoder, a linear layer, and a GAN-based vocoder operating on Mel-spectrograms. LipSound2 \cite{qu2022lipsound2} investigates cross-modal self-supervised pre-training of an encoder-decoder architecture with a location-aware attention mechanism to map face image sequences to mel-scale spectrograms. In contrast, our proposed \ourmodel~differs by using content-rich SSL representations \cite{polyak2021speech} and learning target speech representations from lip sequences. 


\subsection{Self-supervised representation for \lts~synthesis}
Traditional works on \lts~synthesis encode lip or face sequences to hidden states, followed by a decoder to generate Mel-spectrograms. An independently trained vocoder transforms them into time-domain waves. However, these Mel frames are highly correlated along both time and frequency axes which may degrade the performance of entire \lts~synthesis \cite{du2022vqtts}. Moreover, these Mel-spectrograms have higher variance than that of quantized speech SSL representations increasing the complexity of training a \sts~model. Due to its recent emergence, the utilization of SSL representation in \lts~synthesis remains limited. 

In VCVTS \cite{wang2022vcvts}, vector quantized contrastive predictive coding (VQCPC) units are extracted from lip movements, and a speaker encoder, pitch predictor, and decoder are used to infer Mel frames. A separate voice conversion model and vocoder are employed for speaker representation learning and speech synthesis. 

Revise \cite{hsu2023revise} employs a combined audio-visual speech recognition module (P-AVSR - initialized with AV-HuBERT - an audio-visual SSL model) and a modified text-to-speech synthesis module (P-TTS) for generalized speech enhancement tasks. P-AVSR predicts discrete units derived from a self-supervised speech model, and P-TTS converts these units into speech using a modified HiFi-GAN\cite{polyak2021speech} trained on the LJSpeech\cite{ljspeech17} dataset.
Our work closely relates with \cite{hsu2023revise} utilizing SSL representations for \lts~synthesis. However, we utilize disentangled AV-HuBERT and HuBERT \cite{hsu2021hubert} features and a decoupled training procedure to train a \sts~model. Additionally, the existing works have not yet fully capitalized on SSL representations for speaker-specific \lts~generation, and our efforts aim to address this gap. 



\section{Method}
\subsection{Preliminaries}
\ourmodel~consists of three modules: an encoder that extracts lip and speech representation from their corresponding sequences, a \sts~model that maps lip representations to speech representations, and a vocoder that synthesize speech using the speech representations. We introduce four functions as follows:
\begin{itemize}
    \item $f_\text{l}:L^{T\times W \times H} \mapsto L_{\text{ssl}}$, which maps the input lip sequence to its corresponding SSL representation. Here, $T$ represents the number of time-steps (frames), and $H$ and $W$ correspond to the spatial dimensions of the frames.

    \item $f_\text{s}:X \mapsto S_{\text{ssl}}$, which maps the ground-truth raw speech to its corresponding SSL representation.

    \item $f_{\text{s2s}}:L_{\text{ssl}} \mapsto S_{\text{ssl}}$, which maps the lip representation to its corresponding speech representation.

    \item $f_{\text{voc}}:S_{\text{ssl}} \mapsto \widehat{X}$, which maps the speech representation to the synthesized speech $\widehat{X}$.
\end{itemize}
    




\subsection{Encoder}
Although our framework is compatible with various off-the-shelf SSL models, we specifically utilize AV-HuBERT \cite{shi2022learning} for $f_\text{l}$, our video encoder to extract lip representations. We use HuBERT \cite{hsu2021hubert} for $f_\text{s}$, to extract speech representation of target speech signal. The HuBERT and AV-HuBERT models, are trained using a masked-prediction loss to predict cluster IDs, which are learned using k-means clustering. They are initialized with MFCC features derived from acoustic frames, and subsequently more complex features derived from an audio or audio-visual encoder, depending on the specific model being used.
AV-HuBERT incorporates a modified ResNet \cite{he2016deep,petridis2018audio} as its frontend, coupled with a transformer encoder. 
We opted for AV-HuBERT instead of Visual-HuBERT for this task because AV-HuBERT exhibits superior performance in the lip-reading task, as demonstrated by \cite{shi2022learning}. This improvement is observed when the target cluster IDs are derived from both audio and visual modalities, as opposed to a single modality, whether it be audio or visual.
We also finetune the pretrained AV-HuBERT using an attention-based \sts~cross-entropy loss as in \cite{shi2022learning}.  

\subsection{\sts~model}

In recent years, \sts~models have gained significant attention in the field of cross-domain generation. The core concept of our approach is to align representations from two different domains - visual and audio, that share a common generating process. By recovering correspondences between these domains, we facilitate the transfer of knowledge from one domain to the other. 

Our \sts~model, denoted as $f_{\text{s2s}}$, adopts a \nar-based encoder-decoder architecture to map lip representations to their corresponding speech representations. The encoder and decoder consist of feed-forward transformer blocks with self-attention \cite{vaswani2017attention}, along with $1$-dimensional convolutions inspired by Fastspeech2 \cite{ren2020fastspeech}. A transposed convolution layer is used at the encoder to match the rate of video and audio representations. The encoder takes the lip representation $L_{\text{ssl}}$ and encodes it into a sequence of fixed-dimensional vectors. The decoder generates predictions for all representations of $S_{\text{ssl}}$ simultaneously. We train three versions of the \sts~model:
\begin{itemize}
    \item $f_{\text{s2s-units}}$: This encoder-decoder architecture utilizes Cross-Entropy (CE) loss to train the model on the decoded speech units. The CE loss measures the difference between the decoded units and the ground-truth speech HuBERT units. The input to the architecture consists of cluster IDs from the video encoder, and the decoder predicts the corresponding HuBERT cluster IDs for the audio. The objective can be written as:
    \begin{equation}
    \mathcal{L}_{\text{CE}} = -\sum_{i=1}^{N} S_{\text{ssl\_units\_i}} \log(\hat{S}_{\text{ssl\_units\_i}}),
    \end{equation}
    where $S_{\text{ssl\_units}}$ are the ground-truth speech units, $\hat{S}_{\text{ssl\_units}}$ are the decoded speech units, and N are the number of HuBERT units.

    \item $f_{\text{s2s-features}}$: Here the model learns mapping from audio-visual feature to corresponding speech feature vectors. This model utilizes L1 loss, quantifying the difference between the decoded features and ground-truth speech features. The objective can be written as: 
    \begin{equation}
    \mathcal{L}_{\text{L1}} = \frac{1}{T} \sum_{i=1}^{T} |S_{\text{ssl\_features\_i}} - \hat{S}_{\text{ssl\_features\_i}}|,
    \end{equation}
    where $S_{\text{ssl\_features}}$ are the ground-truth speech features, $\hat{S}_{\text{ssl\_features}}$ are the decoded speech features, and $T$ is the time-steps.

    \item $f_{\text{s2s-features-ctc}}$: This architecture follows the same structure as $f_{\text{s2s-features}}$, but also includes an additional fully connected linear head to predict CTC tokens after the encoder layer. For given input lip representation $L_{\text{ssl}} \in \mathbb{R}^{TxD}$ of length T and dimension D, let $Enc_{\text{ssl}}$ be the output of encoder. The goal is to minimize the negative log-likelihood by using $P_{CTC}(S_{\text{ssl}}|Enc_{\text{ssl}})$ to train the model effectively using the CTC approach and is defined as:

    \begin{equation}
        \mathcal{L}_{CTC} := -\log P_{CTC}(S_{\text{ssl}}|Enc_{\text{ssl}}).
    \end{equation}

    By weighted summing the L1 and CTC loss functions, the objective function can be formulated as:
    \begin{equation}
    \label{eq:total-loss-ctc}
    \mathcal{L}_{\text{Tot}} = \alpha_{CTC} * \mathcal{L}_{\text{CTC}} + \alpha_{L1} * \mathcal{L}_{\text{L1}},
    \end{equation}
    where $\alpha_{\text{CTC}} \in \mathbb{R}$ and $\alpha_{\text{L1}} \in \mathbb{R}$ are the hyperparameter that balances the influence between two loss. 


    



\end{itemize}

\subsection{Speech Vocoder}
We use a modified version of HiFiGAN-v2~\cite{kong2020hifi} to synthesize speech. It has a generator $G$ and a discriminator $D$. $G$ runs~$S_{ssl}$ through transposed convolutions for upsampling to recover the original sampling rate followed by residual block with dilations to increase the receptive field to synthesize the signal, $\widehat{X} := G(S_{\text{ssl}})$.

The discriminator in our model has the task of distinguishing the synthesized signal $\widehat{X}$ from the original signal $X$. It is evaluated using two sets of discriminator networks. The multi-period discriminators operate on equally spaced samples of the signals, focusing on capturing temporal patterns and characteristics. On the other hand, the multi-scale discriminators analyze the input signal at different scales, enabling the model to capture both fine-grained details and global structure. The primary objective of the model is to minimize the discrepancy, measured by $D(X,\widehat{X})$, between the original signal and the synthesized signal. This optimization process applies to all the parameters of the speech decoder, improving its overall performance and fidelity.

\section{Experiments}
\subsection{Datasets}
\subsubsection{Lip2Wav} The Lip2Wav dataset \cite{prajwal2020learning} is a large, person-specific, unconstrained dataset, commonly used for learning \lts~synthesis for individual speakers. 
It consists of real-world lecture videos featuring $5$ different speakers. Each speaker has approximately $20$ hours of video data, and the vocabulary size exceeds $5000$ words for each speaker. We do experiments on all five speakers: Chess Analysis (chess), Chemistry Lectures (chem), Hardware Security (hs), Deep Learning (dl), and Ethical Hacking (eh). 

\subsubsection{GRID-4S} The GRID-4S is a subset of the GRID audio-visual dataset \cite{cooke2006audio} specifically designed for constrained \lts~synthesis. This subset includes two male speakers ($s1$, $s2$) and two female speakers ($s4$, $s29$), which are frequently used in the literature \cite{kim2021lip,prajwal2020learning}. The videos in the dataset were captured in an artificial environment. The vocabulary used in GRID-4S is limited to only $51$ words. The sentences in the dataset follow a restricted grammar, with each sentence containing $6$ to $10$ words. 

\subsubsection{TCD-TIMIT-3S} The TCD-TIMIT-3S is a subset of the TCD-TIMIT dataset, which comprises recordings of $62$ speakers captured under studio conditions. Among these speakers, three are trained lip-speakers. The primary objective of selecting this subset was to enable comparison with previous studies \cite{kim2021lip,prajwal2020learning}. Our focus was solely on the audio-visual data generated by these three lip-speakers. Each lip-speaker delivers $375$ distinct sentences that exhibit phonetic diversity. Additionally, there are two sentences that are spoken by all three lip-speakers. 

\subsection{Implementation details}
\subsubsection{Data preparation} For the GRID-4S and TCD-TIMIT-3S datasets, we adhere to the convention of randomly selecting $90\%$ data for training, $5\%$ for validation, and $5\%$ for testing, as established in previous works \cite{michelsanti2020vocoder,prajwal2020learning,vougioukas2019video,wang2022fastlts}. For Lip2Wav, we adopt the official data split \cite{prajwal2020learning}. For consistency, in line with previous works, we evaluated \ourmodel~on the Lip2Wav dataset using a speaker-dependent setting \cite{prajwal2020learning,ephrat2017improved,hong2021speech}. This involved training the network separately with individual speakers. However, for the GRID-4S and TCD-TIMIT-3S datasets, we evaluated \ourmodel~in a constrained (seen) speaker setting \cite{michelsanti2020vocoder,prajwal2020learning,vougioukas2019video}. In this case, we trained a single speaker model for each dataset. The video sequences are resampled to a frame rate of $25$ frames per second (fps), while the raw audio is sampled at $16kHz$. We utilize the SFD \cite{bulat2017far} face detector to detect $68$ keypoints, allowing us to crop a mouth-centered region-of-interest measuring $96\times96$ pixels. In order to solely assess the advantages of using SSL representation in our proposed setup, we opt not to employ any data augmentation techniques to enhance the quality of synthesized speech. The Lip2Wav dataset does not provide transcripts, so we rely on the Whisper \textit{small} model \cite{radford2022robust} to extract transcripts. These transcripts are then used for finetuning the AV-HuBERT model.

\subsubsection{SSL representation} We utilize the official fairseq repository implementation of the BASE models AV-HuBERT \cite{shi2022learning} and HuBERT \cite{hsu2021hubert} for our experiments. We fine-tune the AV-HuBERT pretrained model with an attention-based STS cross-entropy loss for visual speech recognition \cite{shi2022learning}. To achieve this, a transformer decoder is added to the pretrained model, which autoregressively decodes the AV-HuBERT features to target character probabilities. The fine-tuned AV-HuBERT model extracts SSL representations for lip sequences, while HuBERT is employed to extract representations from speech signals. Both models provide $768$-dimensional features. For $f_{\text{s2s-units}}$ model, following the approach in \cite{shi2022learning,hsu2021hubert}, the lip features are clustered into $2000$ AV-HuBERT units, while the speech features are clustered into $100$ HuBERT units, using k-means clustering. For $f_{\text{s2s-features}}$ model, the output features from HuBERT and AV-HuBERT models are used directly.

\subsubsection{\sts~model} Our model comprises a $6$-layer transformer encoder and decoder with a hidden dimension of $512$ and $2$ attention heads. we set the batch size to $32$ and the maximum number of steps to $20,000$. We employ the Adam optimizer with an initial learning rate of $4.4$ x $10^{-2}$, along with an annealing rate of $0.3$ and annealing steps at $[3000, 4000, 5000]$. The HuBERT model encodes speech into features at a frame rate of $50Hz$, while the SSL unit from AV-HuBERT is encoded at $25Hz$. To match these rates, we incorporate a lightweight transposed convolution layer
with a kernel size of $[4, 3]$ and a stride length of $[2, 1]$. We set $\alpha_{\text{CTC}}$ and $\alpha_{\text{L1}}$ mentioned in (\ref{eq:total-loss-ctc}) to $0.001$ and $1$, respectively.
    
\subsubsection{Speech Vocoder}
We use the official implementation of the adapted HiFiGAN-v2\footnote{https://github.com/facebookresearch/speech-resynthesis} to generate audio from speech SSL representations. 
This model employs encoding of raw audio into a sequence of discrete tokens from a set of $100$ possible HuBERT tokens, with a code hop size of $160$ raw audio samples. 
We set the batch size to $16$, the learning rate to $2$x$10^{-4}$, the number of embeddings to $100$, the embedding dimension to $128$, and the model input dimension to $256$. Following the approach in \cite{lee2021direct}, F0 is not used as a feature in our training process. The aforementioned vocoder configuration is effective for speech units. However, for our investigated feature-based models, $f_{\text{s2s-features}}$ and $f_{\text{s2s-features-ctc}}$, we apply a pre-trained k-means \footnote{https://github.com/facebookresearch/fairseq/tree/main/examples/\\textless\_nlp/gslm/speech2unit} clustering model for speech HuBERT units. During the inference phase, the generated features undergo k-means clustering to obtain discrete speech units, which are then

\begin{table}[tbp]
    \centering
    \caption{Performance comparison: \sts~model vs. evaluated variations vs. no \sts~model employed on chemistry speaker of Lip2Wav dataset.}
    \label{tab:sts_evaluations}
    \begin{tabular}{llccc}
    \hline
    \textbf{Baseline (Ours)} & \textbf{STOI} $\uparrow$ & \textbf{ESTOI} $\uparrow$ \\
    \hline
    $f_l\text{(pre-trained)}$ + $f_{voc}$ & 0.447 & 0.22\\
    $f_l\text{(finetuned)}$ + $f_{voc}$ & 0.50 & 0.27 \\
    $f_l\text{(finetuned)}$ + $f_{\text{s2s-units}}$ + $f_{voc}$ & 0.18 & 0.013 \\
    \textbf{$f_l\text{(finetuned)}$ + $f_{\text{s2s-features}}$ + $f_{voc}$} & \textbf{0.583} & \textbf{0.397} \\
    $f_l\text{(finetuned)}$ + $f_{\text{s2s-features-ctc}}$ + $f_{voc}$ & 0.557 & 0.368 \\  
    \hline
    \end{tabular}
    \end{table}

passed through the speech vocoder. We train the vocoders for the three datasets separately. samples.
    
\subsection{Evaluation metric}
During our evaluation, we employ several metrics to assess the quality of the synthesized speech. These include: Word Error Rate (WER), Short-Time Objective Intelligibility (STOI) \cite{taal2011algorithm} and Extended Short-Time Objective Intelligibility (ESTOI) \cite{jensen2016algorithm}. Additionally, we conduct subjective evaluations using Mean Opinion Score (MOS), where human evaluators rate the quality, intelligibility and naturalness of the synthesized speech based on their subjective perception.

\section{Results}
\subsection{Need for \sts~model}
We tested our hypothesis of using a \sts~model on the Lip2Wav dataset, specifically for a chemistry speaker and report our findings in Table~\ref{tab:sts_evaluations}. Fine-tuning the AV-HuBERT model using transcripts consistently improved objective metrics by approximately $0.05$ units on both the metrics compared to the pretrained version. Deploying the \sts~model on the finetuned AV-HuBERT features ($f_l\text{(finetuned)}$ + $f_{\text{s2s-features}}$ + $f_{voc}$) resulted in an increase of approximately $0.08$ and $0.12$ units in STOI and ESTOI metrics, respectively, compared to not using the \sts~model ($f_l\text{(finetuned)}$ + $f_{voc}$). These results highlight the effectiveness of our \sts~approach using SSL representations for \lts~synthesis. The significant performance gap (approximately $0.39$ units on both metrics) between the \sts~model using SSL features and the model using SSL units on both evaluated metrics approximates the amount of information lost in speech reconstruction when audio-visual sequences are represented as SSL units instead of SSL features. From now on, we will refer to $f_l\text{(finetuned)}$ + $f_{\text{s2s-features}}$ + $f_{voc}$ as \ourmodel. The inclusion of CTC loss in our $f_l\text{(finetuned)}$ + $f_{\text{s2s-features-ctc}}$ model resulted in a statistically insignificant decrease of approximately $0.02$ units in STOI compared to the model without CTC loss, $f_l\text{(finetuned)}$ + $f_{\text{s2s-features}}$. The decrease may be due to the lack of ground-truth transcripts in the Lip2Wav dataset. However, when evaluating our model using CTC loss on datasets (GRID-4S and TCD-TIMIT-3S) with ground-truth transcripts, we observed a slight increase of $0.02$ units. Nevertheless, our focus is on working with datasets in the wild that generally lack ground-truth transcripts, so we proceed with experiments excluding the CTC loss. 

\begin{table}[t]
    \centering
    \caption{Performance comparison in constrained-speaker setting on GRID-4S dataset}
    \label{tab:grid}
    \begin{tabular}{llccc}
    \hline
    \textbf{Method} & \textbf{STOI} $\uparrow$ & \textbf{ESTOI} $\uparrow$ & \textbf{WER} $\downarrow$ \\
    \hline
    Vid2speech \cite{ephrat2017vid2speech} & 0.491 & 0.335 & 44.92 \% \\
    Lip2AudSpec \cite{akbari2018lip2audspec} & 0.513 & 0.352 & 32.51 \% \\
    1D GAN-based \cite{vougioukas2019video} & 0.564 & 0.361 & 26.64 \% \\
    Vocoder-based \cite{michelsanti2020vocoder} & 0.648 & 0.455 & 23.33 \% \\
    Ephrat \textit{et al.} \cite{ephrat2017improved} & 0.659 & 0.376 & 27.83 \% \\
    Lip2Wav \cite{prajwal2020learning} & 0.731 & 0.535 & 14.08 \% \\
    VAE-based \cite{yadav2021speech} & 0.724 & 0.540 & - \\
    VCA-GAN \cite{kim2021lip} & 0.724 & \textbf{0.609} & 12.25 \% \\
    kim \textit{et al.} \cite{hong2021speech,kim2021multi} & 0.738 & 0.579 & - \\
    \textbf{\ourmodel} & \textbf{0.754} & 0.571 & \textbf{11.21} \% \\
    \hline
    \end{tabular}
    \end{table}

\begin{table}[t]
    \centering
    \caption{Performance comparison in constrained-speaker setting on TCD-TIMIT-3S dataset}
    \label{tab:tcdtimit}
    \begin{tabular}{llccc}
    \hline
    \textbf{Method} & \textbf{STOI} $\uparrow$ & \textbf{ESTOI} $\uparrow$ & \textbf{WER} $\downarrow$ \\
    \hline
    Vid2speech \cite{ephrat2017vid2speech} & 0.451 & 0.298 & 75.52 \% \\
    Lip2AudSpec \cite{akbari2018lip2audspec} & 0.450 & 0.316 & 61.86  \% \\
    1D GAN-based \cite{vougioukas2019video} & 0.511 & 0.321 & 49.13  \% \\
    Ephrat \textit{et al.} \cite{ephrat2017improved} & 0.487 & 0.310 & 53.52 \% \\
    Lip2Wav \cite{prajwal2020learning} & 0.558 & 0.365 & 31.26 \% \\
    VCA-GAN \cite{kim2021lip} & 0.584 & 0.401 &  - \\
    \textbf{\ourmodel} & \textbf{0.596} & \textbf{0.452} & \textbf{29.03} \% \\
    \hline
    \end{tabular}
    \end{table}
    
\subsection{\ourmodel~in Constrained settings} Table \ref{tab:grid} and \ref{tab:tcdtimit} summarizes the performance of our \ourmodel~in the context of \lts~synthesis using constrained datasets: GRID-4S and TCD-TIMIT-3S. We compare our results with existing \lts~synthesis works, including state-of-the-art approaches. We report the mean test scores on all four speakers of the GRID-4S dataset and all three speakers of the TCD-TIMIT-3S dataset, as documented in previous works. Remarkably, our \ourmodel~approach demonstrates significant improvements in terms of STOI and WER metrics when compared to other approaches. This improvement is particularly noticeable on the TCD-TIMIT-3S dataset, which contains a larger number of novel words that were unseen during training. This observation highlights the ability of our \ourmodel~to accurately pronounce new words and effectively capture semantic information from lip movements, resulting in the generation of more intelligible speech.

\begin{table}[ht]
    \centering
    \caption{Performance comparison in speaker-dependent setting on Lip2Wav dataset}
    \label{tab:lip2wav}
    \begin{tabular}{llccc}
    \hline
    \textbf{Speaker} & \textbf{Method} & \textbf{STOI} $\uparrow$ & \textbf{ESTOI} $\uparrow$ \\
    \hline
    \multirow{5}{4em}{\centering Chemistry Lectures (chem)} & Ephrat \textit{et al.} \cite{prajwal2020learning} & 0.165 & 0.087 \\
    & GAN-based \cite{burnham2013hearing} & 0.192 & 0.132 \\
    & Lip2Wav \cite{prajwal2020learning} & 0.416 & 0.284 \\
    & Hong et al. \cite{hong2021speech} & 0.566 & \textbf{0.429} \\
    & \textbf{\ourmodel} & \textbf{0.583} & 0.397\\
    \hline
    \multirow{5}{4em}{\centering Chess Analysis (chess)} & Ephrat \textit{et al.} \cite{prajwal2020learning} & 0.184 & 0.098 \\
    & GAN-based \cite{burnham2013hearing} & 0.195 & 0.104 \\
    & Lip2Wav \cite{prajwal2020learning} & 0.418 & 0.290 \\
    & Hong et al. \cite{hong2021speech} & 0.506 & 0.334\\
    & \textbf{\ourmodel} & \textbf{0.517} & \textbf{0.340}\\
    \hline
    \multirow{5}{4em}{\centering Deep Learning (dl)} & Ephrat \textit{et al.} \cite{prajwal2020learning} & 0.112 & 0.043 \\
    & GAN-based \cite{burnham2013hearing} & 0.144 & 0.070 \\
    & Lip2Wav \cite{prajwal2020learning} & 0.282 & 0.183 \\
    & Hong et al. \cite{hong2021speech} & 0.576 & 0.402 \\
    & \textbf{\ourmodel} & \textbf{0.627} & \textbf{0.419}\\
    \hline
    \multirow{5}{4em}{\centering Hardware Security (hs)} & Ephrat \textit{et al.} \cite{prajwal2020learning} & 0.192 & 0.064 \\
    & GAN-based \cite{burnham2013hearing} & 0.251 & 0.110 \\
    & Lip2Wav \cite{prajwal2020learning} & 0.446 & 0.311 \\
    & Hong et al. \cite{hong2021speech} & 0.504 & 0.337 \\
    & \textbf{\ourmodel} & \textbf{0.511} & \textbf{0.337} \\
    \hline
    \multirow{5}{4em}{\centering Ethical Hacking (eh)} & Ephrat \textit{et al.} \cite{prajwal2020learning} & 0.143 & 0.064 \\
    & GAN-based \cite{burnham2013hearing} & 0.171 & 0.089 \\
    & Lip2Wav \cite{prajwal2020learning} & 0.369 & 0.220 \\
    & Hong et al. \cite{hong2021speech} & 0.463 & \textbf{0.304} \\
    & \textbf{\ourmodel} & \textbf{0.493} & 0.277\\
    \hline
    \end{tabular}
    \end{table}


\begin{figure}[t]
  \centering
\includegraphics[width=\linewidth]{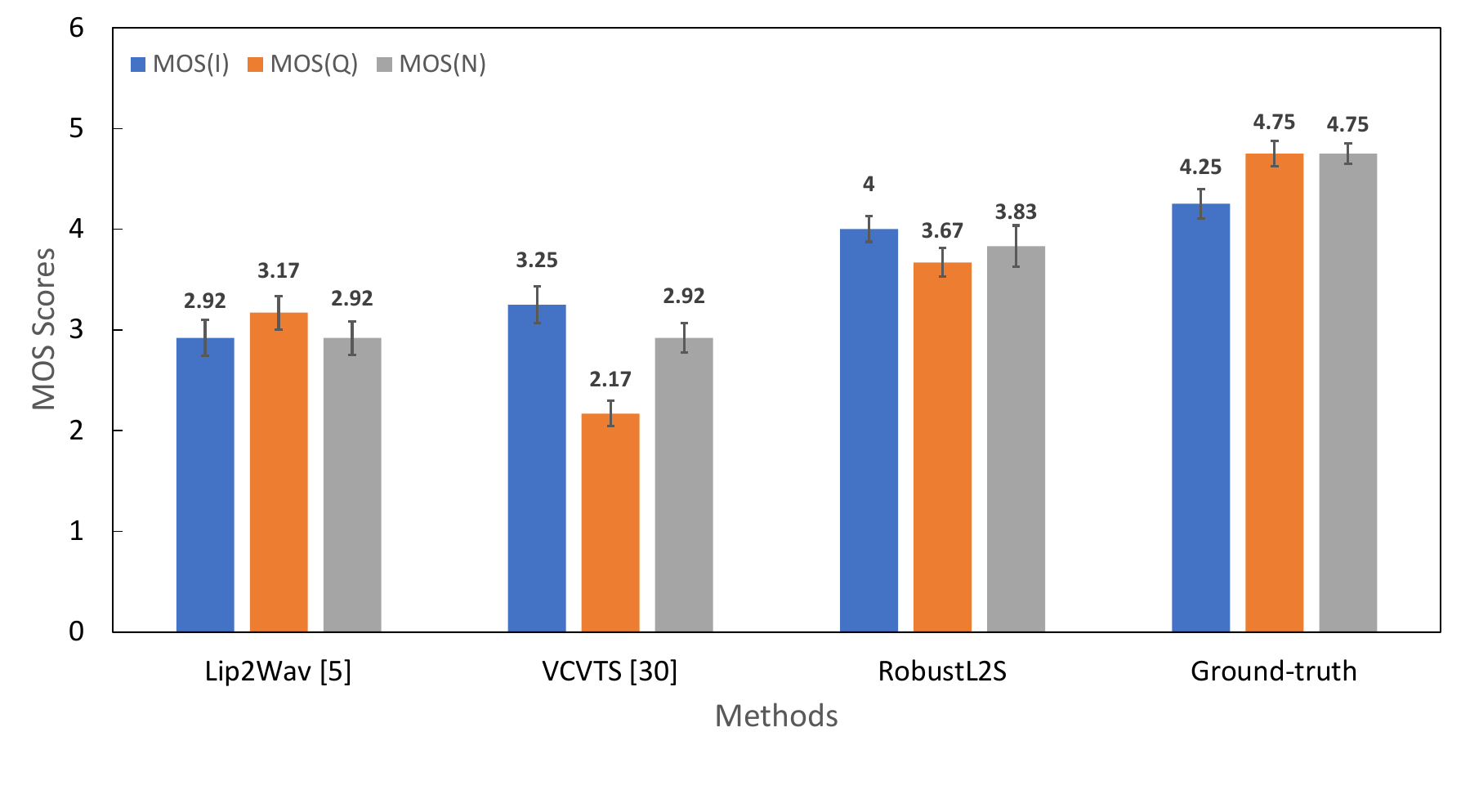}
\caption{MOS scores on Intelligibility, Quality and Naturalness with their 95 \% confidence interval computed from their t-distribution on GRID-4S dataset} 
  \label{fig:mosgrid}
\end{figure}

\begin{figure}[t]
  \centering
\includegraphics[width=\linewidth]{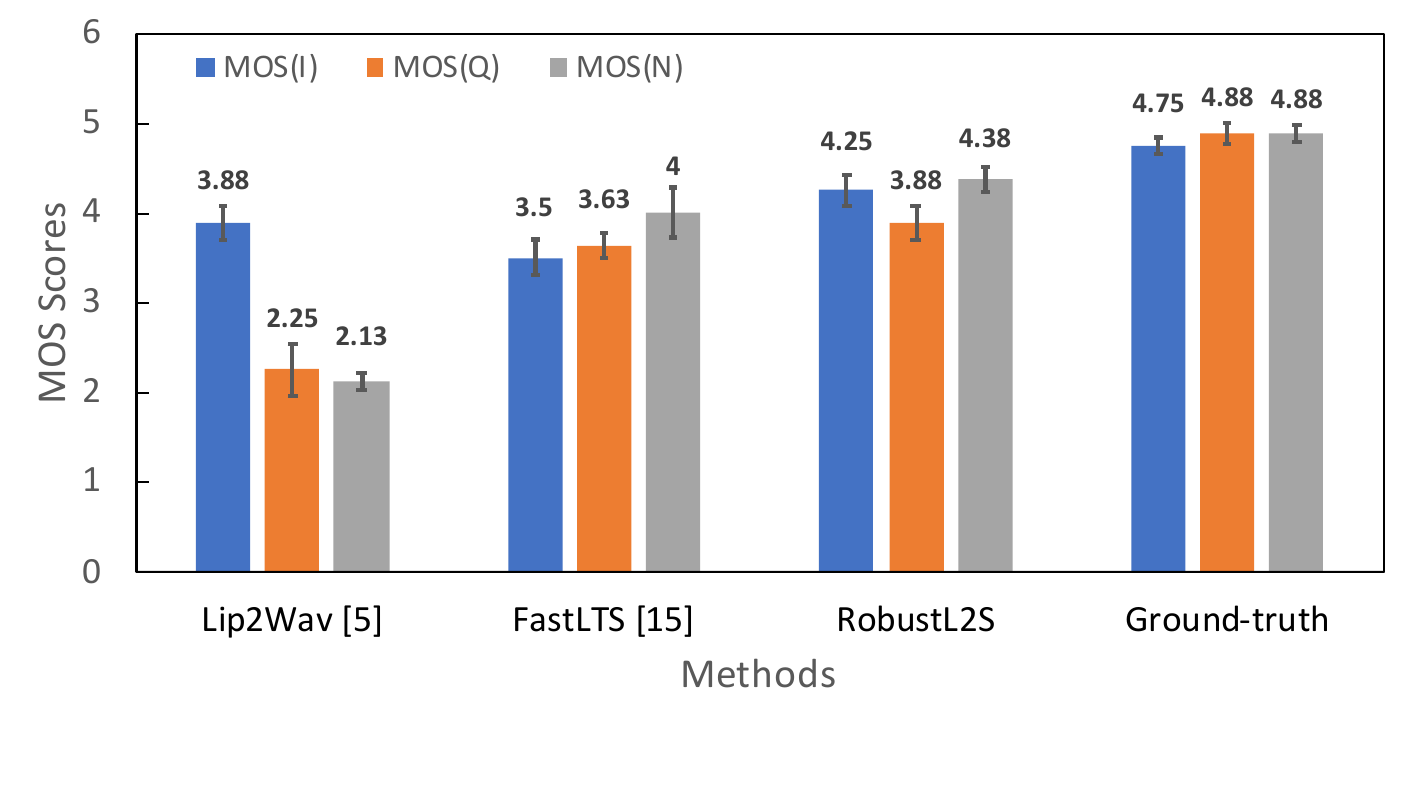}
\caption{MOS scores on Intelligibility, Quality and Naturalness with their 95 \% confidence interval computed from their t-distribution on Lip2Wav dataset} 
  \label{fig:moslip2wav}
\end{figure}

\subsection{\ourmodel~in Unconstrained settings} Table \ref{tab:lip2wav} provides a synopsis of \ourmodel's~performance on the Lip2Wav dataset. This dataset includes a significant amount of silences between words, and \ourmodel~shows a notable improvement across all metrics. Despite Lip2Wav's data asynchrony issues, which may affect the quality of the generated speech, \ourmodel~demonstrates substantial performance gains in objective metrics, highlighting its overall superiority in producing intelligible speech. However, it is worth noting that \ourmodel~performs similarly or slightly worse than \cite{hong2021speech} on the \textit{hs} and \textit{eh} ESTOI metric. This could potentially be attributed to the poor resolution ($480p$ and $360p$) of the original videos, making it challenging to accurately recognize lip regions.

\subsection{Subjective evaluation} Fig. \ref{fig:mosgrid} and Fig. \ref{fig:moslip2wav} shows the MOS scores on intelligibility (MOS(I)), quality (MOS(Q)), and naturalness (MOS(N)) of synthesized speech from evaluated methods on Grid-4S and Lip2Wav datasets. We requested ten English proficient subjects to score five randomly selected samples from different methods on the Lip2Wav and GRID-4S datasets. It can be observed that our model outperforms the evaluated methods, exhibiting higher Mean Opinion Score (MOS) values. This demonstrates that the proposed approach inherits the advantages of disentangled SSL features and the mapping of lip sequences to content-specific information. As a result, our model not only inherently improves the intelligibility aspect of synthesized speech but also generates speech that is highly natural and of high quality.

\section{Conclusions}
We propose a novel framework for \lts~system, called \ourmodel~, which accurately synthesizes spoken content from silent videos. This is accomplished by utilizing a \nar~based sequence-to-sequence model to establish an inter-modality mapping, allowing us to learn a suitable decoding space from the lips' self-supervised (SSL) representations. We further demonstrate the effectiveness of mapping SSL features rather than SSL units for synthesizing intelligible speech. Both quantitative and qualitative results showcase state-of-the-art performance in constrained settings (such as GRID and TCD-TIMIT) and unconstrained settings (like Lip2Wav). In our future work, we aim to introduce emotive effects in the synthesized speech, considering that HuBERT embeddings are known to lack prosody information. Additionally, we plan to explore diffusion-based speech vocoders and their application in a multi-lingual setup. We commit to open-source our implementation on acceptance.


\printbibliography

\end{document}